\documentclass[
preprint,
nofootinbib,
 amsmath,amssymb,
 aps,
]{revtex4-2}
\usepackage{braket}
\usepackage{hyperref}
\usepackage{array}
\usepackage{xcolor}
\usepackage{graphicx}
\usepackage{dcolumn}
\usepackage{bm}

\begin{document}

\preprint{APS/123-QED}

\title{On Inflation and Reheating Features in the  Higgs-$R^2$ Model}

\author{Norma Sidik Risdianto}
\email{norma.risdianto@uin-suka.ac.id}

\affiliation{%
 Department of Physics Education, Universitas Islam Negeri Sunan Kalijaga\\
 Jl. Marsda Adisucipto 55281, Yogyakarta, Indonesia
}%
\date{July 12, 2022}

\begin{abstract}
We investigated inflation in the Higgs-$R^2$ model under the single-field approximation, referred to as the minimal two-field mode, and used this approximation to constrain the Higgs non-minimal coupling $\xi$. During inflation, we examined the impact of $\xi$ on non-Gaussianity and found that it cannot generate a large signal. 
In our paper, the preheating stage is divided into quadratic and quartic regimes. Gauge boson production dominates in the quadratic regime but is insufficient to deplete the inflaton energy. To account for the remaining energy, we introduced a dark matter candidate with a large coupling capable of draining the inflaton's energy density. For $\xi \lesssim 4.2$, the quadratic oscillations extend into the quartic regime, yielding a reheating temperature of order $\sim 10^{10}\,\text{GeV}$, which depends sensitively on the residual inflaton energy density. By contrast, for $\xi > 4.2$, oscillations only happens at the quadratic regime. During this regime, Pauli blocking suppresses the instantaneous decay of gauge bosons into fermions, resulting in a lower reheating temperature of order $\sim 10^9\,\text{GeV}$.

\end{abstract}

\maketitle


\section{Introduction}
Among the various inflationary models, Starobinsky inflation \cite{starobinsky1980new} and Higgs inflation \cite{bezrukov2008standard} remain the most promising candidates. Recent Planck results \cite{akrami2020planck} confirm the consistency of these models with observational data, while disfavoring alternatives such as chaotic and power-law inflation. Starobinsky inflation has been extensively studied, relying solely on gravity as the driving mechanism of inflation. Although the model is theoretically subtle \cite{kehagias2014remarks}, its predictions agree remarkably well with current data \cite{akrami2020planck}. The model is based on an action containing an $R^2$ term, often referred to as the $R^2$ model, representing the simplest class of $f(R)$ gravity theories\footnote{See also Refs. \cite{odintsov2019unification,odintsov2019f,nojiri2020f} for $f(R)$ models related to axions, and Ref. \cite{de2016spotting} for studies on deviations from the $R^2$ model.} (see also \cite{de2010f,sotiriou2010f,nojiri2011unified,nojiri2017modified}). 

Higgs inflation \cite{bezrukov2008standard}, in contrast, is economical in the sense that it introduces no particles beyond the Standard Model (SM). Instead, it relies on a non-minimal coupling $\xi$ between the Higgs field and the Ricci scalar. Despite this advantage, Higgs inflation suffers from a unitarity problem \cite{barbon2009naturalness,burgess2010higgs,hertzberg2010inflation,nozari2024higgs} due to the large value of the coupling, $\xi \sim 10^5$.\footnote{Even independently, a large $\xi$ is considered problematic \cite{gorbunov2019scalaron}.}. On the other hand, the Higgs inflation also suffered from the large $\lambda$ (due to large $\xi$), which should be addressed carefully \cite{nozari2018lowering}. Meanwhile, pure $R^2$ inflation requires an additional field to fully realize the preheating and reheating dynamics \cite{cheong2020beyond}. For further discussion, Refs. \cite{bezrukov2012distinguishing} and \cite{ketov2020equivalence} provide insights into distinguishing between the two models as well as their possible equivalence.

Several studies have been done to settle the problems of both  Higgs and $R^2$ and inflation, for example, to combine them in Higgs-$R^2$ (or generally a Scalar-$R^2$) inflation (see, e.g., Refs. \cite{rigopoulos2006large,byrnes2009non,garcia2020revisiting,battefeld2009non,gundhi2020scalaron,gorbunov2019scalaron,gorbunov2013r2,cheong2020beyond,ghilencea2018two,he2019violent,he2021occurrence,bezrukov2019some,bezrukov2020heatwave,canko2020simple,gialamas2020reheating,salvio2015classical}). The addition of $R^2$ operator may 'heal' the non-minimal coupling of Higgs inflation \cite{gorbunov2019scalaron} also 'taming' the spiky behavior of oscillating inflaton field during preheating \cite{he2019violent}. In the Higgs-$R^2$ inflation, it is found that the cut-off scale is as large as the Planck scale \cite{ema2017higgs}. One of the exciting facts, once Higgs and $R^2$ terms are combined to become Higgs-$R^2$ inflation, pure Higgs inflation can't be obtained, while pure $R^2$ inflation is possible \cite{ema2017higgs, enckell2020higgs}.

Higgs-$R^2$ inflation is expected to exhibit features characteristic of multi-field inflation, including the possibility of generating larger and potentially detectable non-Gaussianity \cite{rigopoulos2006large,byrnes2009non,garcia2020revisiting,battefeld2009non}, in contrast to the typically suppressed signal in single-field models \cite{gangui1993three,maldacena2003non}. Such a distinction could lead to observable signatures, for instance, in data from the Planck satellite \cite{akrami2020planck}, thereby providing insight into whether a single-field or multi-field dynamics drives inflation. 

Refs. \cite{shojaee2018more,nozari2016large} indicate that even single-field inflationary models are capable of generating a suitable level of non-Gaussianity, challenging the common expectation that multiple fields are necessary. Nevertheless, the role of multi-field inflation in enhancing non-Gaussianity remains debated. Ref. \cite{battefeld2009non} demonstrates that, under slow-roll conditions, multi-field models may yield only negligible effects, whereas Ref. \cite{byrnes2009non} shows that significant enhancement arises only in the late stages of inflation, when slow-roll is violated. By contrast, other works, such as Refs. \cite{enqvist2005non,kohri2010generation}, suggest that non-Gaussianity can be substantially amplified during the preheating phase.

The inflationary dynamics of the Higgs-$R^2$ model have been extensively studied in Refs. \cite{rigopoulos2006large,byrnes2009non,garcia2020revisiting,battefeld2009non,gundhi2020scalaron,gorbunov2019scalaron,gorbunov2013r2,calmet2016higgs,ghilencea2018two}. In this work, we focus on the effective single-field trajectory, known as the \textit{minimal two-field mode} \cite{he2019violent}. In this paper, we will constrain the non-minimal coupling $\xi$ of the Higgs operator. We will see whether $\xi$ can provide sufficient non-Gaussianity. On the other hand, the preheating dynamics of the Higgs-$R^2$ model have been investigated in several works (see, e.g., Refs. \cite{he2019violent,he2021occurrence,bezrukov2019some,bezrukov2020heatwave}). For comparison, in single-field inflation with a large non-minimal coupling, preheating is highly violent, leading to an efficient depletion of the inflaton potential energy \cite{ema2017violent}. In contrast, preheating in the Higgs-$R^2$ model proceeds more smoothly, with the inflaton energy drained less effectively \cite{ema2017violent}. Moreover, tachyonic preheating can occur in this model \cite{he2021occurrence,bezrukov2020heatwave}, which in principle enhances the efficiency of energy transfer. However, in the minimal two-field mode, this effect is suppressed.

In inflation with a large non-minimal coupling\footnote{Here, we refer to ``large'' as $\mathcal{O}(10^2)$ or greater, and ``small'' as $\mathcal{O}(10)$ or less.}, the onset of the preheating stage corresponds to the scalaron field value in the Einstein frame $\tilde{\phi}\lesssim M_p$, where the potential can be approximated by a quadratic potential. Many studies treat this quadratic regime as the dominant stage in which the inflaton energy is efficiently depleted. For example, see Ref. \cite{bezrukov2009initial} for Higgs inflation\footnote{See also Refs. \cite{garcia2009preheating,repond2016combined} for lattice studies of Higgs preheating.} and Ref.~\cite{ema2017violent} for the general single-field case, as well as Refs. \cite{he2019violent,he2021occurrence} for the Higgs-$R^2$ model. 
For small non-minimal coupling, however, the quadratic regime can be neglected. In this case, preheating is more appropriately described as an effective quartic potential (see, e.g., \cite{ballesteros2017standard,hashimoto}). In both regimes—quadratic or quartic—the reheating temperature is determined by the perturbative decay of the daughter fields produced during inflaton oscillations. The relative importance of the quadratic and quartic stages depends sensitively on the value of the non-minimal coupling in the single-field case. In this work, we investigate how the non-minimal coupling $\xi$ affects both the preheating dynamics and the resulting reheating temperature. In this paper, we restrict our analysis to $\xi > 1$.

\section{The Higgs-$R^2$ Inflation: Features in The Inflationary Stage}
In this section, we discuss the main features of Higgs-$R^2$ inflation. Since it belongs to the class of multi-field inflation models, we outline its characteristic properties and their implementation in the Higgs-$R^2$ model. We then examine the implications for non-Gaussianity and, in particular, constrain the Higgs non-minimal coupling $\xi$ based on the suppressed non-Gaussianity predicted by the effective single-field inflation.

\subsection{The multi-field inflation and slow-roll parameters}\label{subsectionmultifield}
In this subsection, we will discuss the case of multi-field inflation and then simplify it into the Higgs-$R^2$ model later in the next subsection. We start by writing the action with $d$ number of scalar fields $\psi^a$ ($a=1,2,3,...d$) as
\begin{equation}\label{actionmultifield}
    \begin{split}
        S=\int d^4x\sqrt{-g}\left[ \frac{M^2_p}{2}R-\frac{1}{2}\gamma_{ab}g^{\mu\nu}\partial_\mu \psi^a \partial_\nu \psi^b-U(\psi)\right],\\
    \end{split}
\end{equation}
where $R$ is the Ricci scalar with metric tensor $g_{\mu \nu}$ (and determinant $g$). We also have the field metric $\gamma_{ab}$, corresponding to the mixing kinetic terms. $U(\psi)$ is the potential with the function of $\psi^a$. We also use $M_p=1/\sqrt{8\pi G}$ as the reduced Planck mass. 

To obtain the equation of motions, we take the field to be only time-dependent as $\psi^a(\textbf{x},t)=\psi_0^a(t)$, hence we can write three background equations:
\begin{equation}\label{backgroundequation}
\begin{split}
        &H^2=\frac{1}{3M_p^2}\left[ \frac{1}{2}\Dot{\psi}^2_0+U(\psi)\right],\\
        &\Dot{H}=-\frac{\Dot{\psi}^2_0}{2 M_p^2},\\
        &D_t \Dot{\psi}^a_0+3 H \Dot{\psi}^a_0+U^a=0.\\
\end{split}
\end{equation}
We have defined the derivative potential $U_a=\partial U/\partial \psi^a$ and Hubble parameter $H=\Dot{a}/a$, where $a(t)$ is the scale factor. Throughout this paper, the dot always represents the derivative with respect to physical time. Here we also introduce the covariant derivatives $D\Phi=d\Phi + \Gamma^a_{bc}\Phi^b d\psi^c_0$ and $D_t=D/dt$. Please note, the Christoffel symbols we used are coming from the field metric $\gamma_{ab}$ as $\Gamma^a_{bc}=(1/2)\gamma^{ad}(\partial_b \gamma_{dc}+\partial_c \gamma_{bd}-\partial_d \gamma_{bc})$. In addition, we also used $\Dot{\psi}^2_0=\gamma_{ab}\psi^a_0\psi^b_0$. 

Later, we will define the tangent direction $T^a$ and the normal direction $N^a$ to discuss the trajectory features. Both terms are defined as \cite{achucarro2011features}
\begin{equation}\label{tn}
    \begin{split}
        &T^a\equiv \frac{\Dot{\psi^a_0}}{\Dot{\psi_0}},\\
        &N^a\equiv\text{sign}(t)\left(\gamma_{bc}D_t T^b D_t T^c\right)^{-1/2} D_t T^a,
    \end{split}
\end{equation}
where $\text{sign}(t)=\pm 1$ responsible to adjust the orientation of $N^a$ with respect to $D_t T^a$. Please note, $T^a$ and $N^a$ obey the relation $N^a N_a=T^a T_a=1$ and $N^aT_a=0$. In addition, we can write $D_t T^a$ with the help of Eq. \eqref{backgroundequation} and \eqref{tn}. Hence we obtain
\begin{equation}\label{dt}
    D_t T^a=  -\frac{\Ddot{\psi}_0}{\Dot{\psi}_0}T^a-\frac{1}{\Dot{\psi}_0}\left(3H\Dot{\psi}^a_0+U^a \right).
\end{equation}
Projecting the last equation to $T^a$ and $N^a$ separately, we get the new equations:
\begin{equation}\label{dt2}
        \Ddot{\psi}_0+3H\Dot{\psi}_0+U_\psi=0
\end{equation}
and
\begin{equation}\label{dt3}
        D_tT^a=-\frac{U_N}{\Dot{\psi}_0}N^a,
\end{equation}
where in Eq. \eqref{dt2} and \eqref{dt3} we defined $U_\psi=T^aU_a$ and $U_N=N^aU_a$. In addition, they obey the relation $U_a=U_\psi T^a+U_N N^a$. 

With all requirements fulfilled, we can determine the slow-roll parameters.  Straightforwardly, we can write the slow-roll parameters as
\begin{equation}\label{epsiloneta}
    \epsilon =-\frac{\Dot{H}}{H^2}=\frac{\Dot{\psi}^2_0}{2M_p^2H^2}, \hspace{1cm}
    \eta^a=-\frac{1}{H \Dot{\psi}_0}D_t\Dot{\psi}^a_0.
\end{equation}
Additionally, $\eta^a$ can be decomposed into $\eta^a=\eta^\parallel T^a+\eta^\perp N^a$, thus we obtain two other parameters
\begin{equation}\label{etaparalel}
    \eta^\parallel\equiv -\frac{\Ddot{\psi}_0}{H \Dot{\psi}_0},
\end{equation}
which can be regarded as the usual second slow-roll parameter and
\begin{equation}\label{etaperp}
    \eta^\perp\equiv \frac{U_N}{H \Dot{\psi}_0},
\end{equation}
as the turn parameter, which bends the trajectory of inflaton. We can also have the second turn parameter $\zeta^\perp$ as
\begin{equation}
    \zeta^\perp \equiv -\frac{\Dot{\eta}^\perp}{H\eta^\perp}.
\end{equation}
The last parameter will be redundant for the rest of this paper.

Please note that $\epsilon$ and $\eta^\parallel$ tend to be small during inflation for slow-roll conditions, but $\eta^\perp$ does not have any constraint \cite{achucarro2011features}, it could be large. Thus large bending from the geodesic is also possible. The radius of curvature $\sigma$ can be defined by
\begin{equation}\label{sigma}
\frac{1}{\sigma}=\left(\gamma_{ab} \frac{DT^a}{d\psi_0}\frac{DT^b}{d\psi_0}\right)^{1/2}=\frac{1}{\Dot{\psi}_0}\left(\gamma_{ab} \frac{DT^a}{dt}\frac{DT^b}{dt}\right)^{1/2}=H\eta^\perp,
\end{equation}
where the last line is provided by the help of Eq. \eqref{tn},\eqref{dt3}, and \eqref{etaperp}. Also, using Eq. \eqref{epsiloneta}, we can obtain
\begin{equation}\label{eta1}
    |\eta^{\perp}|=\sqrt{2\epsilon}\frac{M_p}{\sigma \Dot{\psi}_0}.
\end{equation}
We will later use the last result to calculate the non-Gaussianity of multi-field inflation. Furthermore, even though $\eta^\perp$ is not constrained, it is suppressed by the smallness of $\sqrt{\epsilon}$. Hence, the largeness of $\eta^\perp$ strongly depends on $\sigma$ and $\Dot{\psi}$.

\subsection{A short preview of Higgs-$R^2$ inflation}
We start from the following action in the Jordan frame of Higgs-$R^2$ inflation as
\begin{equation}\label{h-r2action}
\begin{split}
      S_J=\int d^4x\sqrt{-g_J}&\Bigg[ \frac{1}{2}M_{p}^2\left(R_J+\frac{R^2_J}{6M^2} \right) +\frac{\xi h^2R_J}{2}-\frac{g^{\mu\nu}_J}{2}\partial_\mu h \partial_\nu h-\frac{1}{4}\lambda h^4\Bigg], \\
\end{split}
\end{equation}
where $g_J$ and $R_J$ are respectively the determinant of the metric $g^{\mu\nu}_J$ and Ricci scalar in the Jordan frame.  Here we take $h$ as the Higgs field in the unitary gauge, $\xi$ as the non-minimal coupling between the Higgs field with gravity, and $\lambda$ as the Higgs quartic coupling. In this paper, we used $\lambda=0.01$.

We can transform the action in Jordan frame  $S_J$ to Einstein frame $S_E$ via Weyl transformation as
\begin{equation}\label{conformal}
\begin{split}
   & g^{\mu\nu}=\Omega g_J^{\mu\nu},\\
   &\Omega\equiv 1+\frac{R_J}{3M^2}+\frac{\xi h^2}{M^2_p}\equiv e^{\sqrt{\frac{2}{3}}\frac{\phi}{M_p}}\equiv e^\chi,\\
\end{split}
\end{equation}
which we defined the scalaron field $\phi$ here \cite{gorbunov2019scalaron,ema2017higgs,bezrukov2019some}. Please note, $M$ in Eq. \eqref{h-r2action} and \eqref{conformal} corresponds to the effective mass of the scalaron during the small field value. After transformation, we should obtain the action in the Einstein frame $S_E$ as
\begin{equation}\label{se}
        S_E=\int d^4x\sqrt{-g}\left[ \frac{M^2_p}{2}R-\frac{1}{2}g^{\mu\nu}\partial_\mu \phi \partial_\nu \phi -\frac{1}{2}e^{-\chi} g^{\mu\nu}\partial_\mu h \partial_\nu h-U(\phi,h)\right],
\end{equation}
with potential
\begin{equation}\label{potential}
    U(\phi,h)= \frac{1}{4}\lambda h^4 e^{-2\chi}+\frac{3}{4}M^2_p M^2\left[1-\left( 1+\frac{\xi h^2}{M_p^2} \right) e^{-\chi}\right]^2.
\end{equation}
Please note, action in Eq. \eqref{se} is similar with action \eqref{actionmultifield} for the Higgs-$R^2$ model with
\begin{equation}
    \gamma_{ab}=\left(\begin{array}{cc}
        1 & 0   \\ 
        0 & e^{-\chi}
    \end{array}\right), \hspace{1cm} \psi^a=\left(\phi, 
        h\right).
\end{equation}

In this work, we assume that in Higgs-$R^2$ inflation the inflaton trajectory follows a single-field approximation, referred to as the \textit{minimal two-field mode}. We assume the inflaton follows the path of the potential's minimum\footnote{This assumption is also used by Ref. \cite{ballesteros2017standard, hashimoto}}. In this regime, the fields satisfy the relation $h^2 = \xi R_J / \lambda$ \cite{ema2017higgs,he2019violent}.
Inserting this result to Eq. \eqref{conformal}, we obtain 
\begin{equation}\label{h2}
    h^2=\frac{e^{\chi}-1}{\frac{\lambda}{\xi M_p^2}\left( \frac{\xi^2}{\lambda}+\frac{M_p^2}{3M^2} \right)}.
\end{equation}
In addition, the potential represents the minimal two-field mode is \cite{ema2017higgs}
\begin{equation}\label{potentialminimum}
    U(\phi)=\frac{M_p^4}{4}\frac{ 1}{\frac{\xi^2}{\lambda}+\frac{M_p^2}{3M^2}}\left(1-e^{-\chi} \right)^2.
\end{equation}
Based on Eq. \eqref{potentialminimum}, we can conclude the potential for a minimal two-field mode dominantly following the scalaron's trajectory with some deviation due to the shifting parameters. Using the last potential on the power scalar spectrum through constraint from the Cosmic Microwave Background (CMB) \cite{akrami2020planck}, we obtain\footnote{The recent results from the Atacama Cosmology Telescope \cite{louis2025atacama} show that, for the Starobinsky model to satisfy the $68\%$ CL, its e-folds $N_f$ must be pushed higher. Typically, a shift in the number of e-folds could significantly affect the reheating temperature. However, since in our paper the reheating temperature is not evaluated through $N_f$, the ACT results may not substantially alter our conclusions.} 
\begin{equation}\label{cmb}
    \frac{\xi^2}{\lambda}+\frac{M_p^2}{3M^2}=\frac{N^2_{f}}{72\pi^2 \mathcal{A}_s}\equiv \frac{1}{\mathcal{C}}\approx 2.1 \times 10^9,
\end{equation}
where we pick $N_f=56$ correspond to the number of e-folds, $\mathcal{A}_s=2.1\times 10^{-9}$ refers to the primordial  spectrum perturbation on the pivot scale $k_o=0.05\text{Mpc}^{-1}$ \cite{akrami2020planck,he2019violent}. Based on these values, the parameter $\xi$ is bounded between $\xi = 1$ (corresponding to nearly pure $R^2$ inflation) and $\xi \lesssim 4500$ (corresponding to nearly pure Higgs inflation). It should be noted that our discussion covers both limiting cases as well as the intermediate range.

With this in mind, Eq. \eqref{h2} can be rewritten in the simpler form
\begin{equation}\label{h2x}
    h^2 = \mathcal{C}\,\frac{\xi M_p^2}{\lambda}\left(e^{\chi}-1\right).
\end{equation}
This expression allows us to work directly with the parameters $\xi$ and $\lambda$, which are intrinsic to Higgs inflation. It is important to note that $\xi^2/\lambda > M_p^2 / (3M^2)$ corresponds to the Higgs-like regime, while $\xi^2/\lambda < M_p^2 / (3M^2)$ corresponds to the $R^2$-like regime \cite{ema2017higgs,he2019violent}. The relation in Eq.~\eqref{h2x} is model-independent and applies to both regimes, provided the minimal two-field mode is preserved. 
As we shall see later, large values of $\xi$ are naturally disfavored during preheating. However, we temporarily set aside the unitarity issue associated with large $\xi$ and instead focus on its dynamical impact in this regime.

One can also analytically solve Eq. \eqref{h2x} during inflation. We can assume $e^\chi\gg 1$, hence we obtain the relation
\begin{equation}\label{h2xx}
    h^2\simeq\mathcal{C}\frac{\xi M_p^2}{\lambda}e^{\chi}.
\end{equation}
During the end of inflation  (when $\epsilon=1$ and $\phi\approx 1 \hspace{1mm}M_p$), the inflaton starts to oscillate. Additionally, the Higgs field has a value
\begin{equation}\label{higgsend}
    \tilde{h}_\text{end}^2\simeq \sqrt{\frac{2}{3}}\mathcal{C}\frac{\xi M_p}{\lambda}\tilde{\phi}_\text{end},
\end{equation}
where the tilde corresponds to the amplitude of the corresponding fields.
Lastly, the critical field value is obtained when
\begin{equation}\label{criticalhiggs}
   \tilde{\phi}_\text{crit}\approx \tilde{h}_\text{crit}\simeq \sqrt{\frac{2}{3}}\mathcal{C}\frac{\xi M_p}{\lambda}.
\end{equation}
The last two equations are only valid if we preserve the minimal two-field mode, which is described shortly in the preheating stage (see section \ref{quadratic} and \ref{quartic}).
Considering the large $\xi$ corresponding to Higgs-like inflation, the critical field value can be pushed higher. For example, if we consider the nearly-pure Higgs inflation $\xi=4500$, the corresponding critical field value of Higgs is $1.75\times 10^{-4}\hspace{1mm}M_p$. However, if we take the small minimal coupling $\xi=10$, we got only $3.88 \times 10^{-7} \hspace{1mm} M_p$. This determination is really important in the late stage of the preheating regime. We will show later that the non-minimal coupling could potentially affect the reheating temperature.

\subsection{The turn parameter in Higgs-$R^2$ inflation.}
In this part, we will explicitly describe the features of Higgs-$R^2$ using the features of general multi-field inflation depicted in \ref{subsectionmultifield}. Straightforwardly, we can obtain the tangent and normal  trajectory, respectively as \cite{he2018inflation}
\begin{equation}\label{T}
    T^a=\frac{\Dot{\psi}^a_0}{\Dot{\psi}_0}=\frac{1}{\sqrt{\Dot{\phi}^2+e^{-\chi}\Dot{h}^2}}\left(\Dot{\phi},\Dot{h}\right),
\end{equation}
\begin{equation}\label{N}
    N^a=\text{sign(t)}\left(\gamma_{bc} \frac{DT^b}{dt}\frac{DT^c}{dt}\right)^{-1/2}\frac{DT^a}{dt}=\frac{e^{\chi/2}}{\sqrt{\Dot{\phi}^2+e^{-\chi}\Dot{h}^2}}\left(-e^{-\chi}\Dot{h},\Dot{\phi} \right),
\end{equation}
where partially can solve \cite{he2018inflation}
\begin{equation}\label{thetadot}
    \left(\gamma_{ab} \frac{DT^a}{dt}\frac{DT^b}{dt}\right)=e^{\chi}\frac{\left(\frac{\partial U}{\partial h}\Dot{\phi}-e^{-\chi}\frac{\partial U}{\partial \phi} \Dot{h}\right)^{2}}{\left(\Dot{\phi}^2+e^{-\chi}\Dot{h}^2\right)^{2}}.
\end{equation}
These tools can approximate the turn parameter $\eta^\perp$ from Eq. \eqref{sigma} and \eqref{thetadot}. Finally, we obtain
\begin{equation}\label{eta2}
    e^{\chi/2}\frac{\left(\frac{\partial U}{\partial h}\Dot{\phi}-e^{-\chi}\frac{\partial U}{\partial \phi} \Dot{h}\right)}{\left(\Dot{\phi}^2+e^{-\chi}\Dot{h}^2\right)}=H{\eta^\perp}.
\end{equation}
Solving the last result may take much effort.
However, even if $\partial U/\partial \phi\gg \partial U/\partial h$, $\partial U/\partial \phi$ is suppressed by $e^{-\chi}$ and $\Dot{h}$ which makes it  several order smaller than $\partial U/\partial h$. Hence, we can obtain the simplified result by using $e^{-\chi}3H\Dot{h}\approx \partial U/\partial h$ (see Eq. \eqref{hdotdot}) as
\begin{equation}
        3e^{-\chi/2}\frac{\Dot{h}}{\Dot{\phi}^2}\approx {\eta^\perp}.
\end{equation}
If we used the ratio of $\Dot{h}/\Dot{\phi}$ in Eq. \eqref{h2xx}, we finally get
\begin{equation}\label{etamin}
    \eta^{\perp}\approx 3 e^{-\chi/2}\frac{ \Dot{h}}{\Dot{\phi}}= 3\sqrt{\frac{2}{3}\frac{\xi}{\lambda} \mathcal{C}}.
\end{equation}
Thus, we obtain the relation on the turn parameter $\eta^\perp$. We found that the turn parameter $\eta^\perp$ strongly depends on $\xi/\lambda$, which are properties of Higgs inflation. 
 
\subsection{The non-Gaussianity in the Higgs-$R^2$ inflation}
In this section, we will use the formalism of non-Gaussianity which is introduced by Ref. \cite{rigopoulos2006large} (see also \cite{rigopoulos2005non} and \cite{rigopoulos2006nonlinear}) for analytical and Ref \cite{rigopoulos2007quantitative} for the numerical calculation. The parameter $f_\text{NL}$ provided by this formalism is
\begin{equation}\label{fnl1}
    f_\text{NL}=f(\epsilon,\eta^\parallel, \eta^\perp,\zeta^\perp, U(\phi,h),  N_f).
\end{equation}
It is better to write Eq. \eqref{fnl1} more explicitly. Before proceeding, we should make an approximation because this case is quite complicated in the original paper in Ref. \cite{rigopoulos2006large}. In this approximation,  $f_\text{NL}$ can be written by
\begin{equation}\label{fnl}
     f_\text{NL}=2(\epsilon +3\eta^\parallel-\zeta^\perp/\eta^\perp)+\Psi N_f,
\end{equation}
where
$\Psi\simeq 4(\eta^\perp)^2$. 

The small non-Gaussianity $f_\text{NL}\ll 1$ may be acquired since it is a function of slow-roll parameters. Thus, from Eq. \eqref{fnl}, the large non-Gaussianity can be expected at the end of inflation when $\epsilon=1$.
In that case, the dominant term for non-Gaussianity during inflation $f_\text{NL}\geq 1$ can be achieved if $4(\eta^\perp)^2  N_f\geq 1$. As for reminder $N_f$
is the number of e-folds, which means it could produce quite a large number. $\eta^\perp$ doesn't need to be large, as long as it is not too small ($\eta^\perp\sim 0.07$ would be sufficient for $N_f \sim 50$). Using these criteria, we can simplify Eq. \eqref{fnl} to be
\begin{equation}
    f_\text{NL}\approx 4(\eta^\perp)^2  N_f= 24 \mathcal{C}^2N_f\hspace{1mm} \frac{\xi}{\lambda},
\end{equation}
where in the last result we have substituted  $\eta^{\perp}$ from Eq. \eqref{etamin}. If we put non-minimal coupling to be nearly-pure Higgs inflation with $\xi= \mathcal{O}(10^3)$ for $\lambda=0.01$, the contribution to the non-Gaussianity is still $\ll 1$. With this result, it is convenient to say that the non-minimal coupling can be regarded as a free parameter since we can take it to be almost any value. Thus, the spectrum of this model can be extended from nearly-pure $R^2$ inflation to nearly-pure Higgs inflation. However, in the next section, we will try to constrain the non-minimal coupling $\xi$ depending on its impact on preheating and reheating.

In the minimal two-field mode of the Higgs-$R^2$ model, it is noteworthy that the parameters of the Higgs sector play the dominant role. In this model, the parameter of $R^2$ inflation, $M$, is effectively absorbed into the constant $\mathcal{C}$ and thus remains hidden.

\section{Preheating in The Quadratic Regime}\label{quadratic}

\subsection{The self-production of the inflaton field}
At the end of inflation, when the field value is $\tilde{\phi}<1\hspace{1mm}M_p$, the inflaton starts to oscillate. At this stage, the inflaton will decay non-perturbatively via parametric resonance. The equations of motion corresponding to the scalaron and Higgs boson can be depicted as

\begin{equation}\label{phidotdot}
    \Ddot{\phi}+3H\Dot{\phi}+\frac{1}{\sqrt{6}M_p}e^{-\chi}\Dot{h}^2+\frac{\partial U}{\partial \phi}=0,
\end{equation}
\begin{equation}\label{hdotdot}
    \Ddot{h}+3H\Dot{h}-\sqrt{\frac{2}{3}}\frac{\Dot{\phi}\Dot{h}}{M_p}+e^{\chi}\frac{\partial U}{\partial h}=0
\end{equation}
and an extra
\begin{equation}\label{hubble2}
    3M^2_p H^2 = \frac{1}{2}\Dot{\phi}^2+\frac{1}{2}e^{-\chi}\Dot{h}^2+ U(\phi, h).
\end{equation}

\begin{figure}[t]
\centering
\includegraphics[width=10cm]{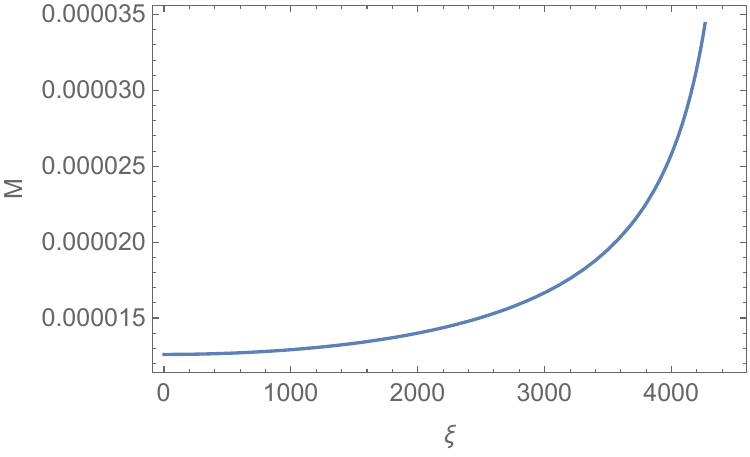}
\caption{Plot $\xi$ with $M$  is in order of Planck mass $M_p$. Both parameters are constrained by CMB result on the scalar power spectrum $\mathcal{A}_s$ and the number of e-fold $N_f$.}
\label{xi-m}
\end{figure}

The numerical calculation corresponding to the last three equations can be seen in \cite{he2019violent,he2021occurrence,bezrukov2020heatwave}. However, we want to simplify the result by assuming the inflaton's trajectory in a single-field approximation\footnote{Our chosen analytical single-field trajectory is invalid if it is nearly pure Higgs inflation. However, such a condition is discouraged by Ref. \cite{ema2017higgs}. Hence, our approximation is valid for the general case.}. With our chosen inflaton's direction, we may lose the complexity of the scalaron and Higgs directions as depicted in Ref. \cite{he2019violent,he2021occurrence,he2021perturbative,bezrukov2020heatwave}, for example, the tachyonic behavior due to Higgs' negative mass and the burst of particle productions due to Higgs direction when $\phi<0$. We expect the preheating case to be calculated semi-analytically in our fine-tuning conditions.
With the selected fine-tuning, the equation of scalaron can be depicted by
\begin{equation}\label{eominflaton}
    \Ddot{\phi}+3H\Dot{\phi}+\frac{\partial U}{\partial \phi}=0,
\end{equation}
the potential is written by the function of $\phi$ (see Eq. \eqref{potentialminimum}) only. Furthermore, we obtain
\begin{equation}\label{tildem}
    U(\phi)=\frac{3}{4}M_p^2\tilde{M}^2\left(1-e^{-\chi} \right)^2, \hspace{0.5cm}\tilde{M}^2=\frac{M^2}{1+\frac{3\xi^2 M^2}{\lambda M_p^2}}=\frac{M_p^2}{3}\mathcal{C},
\end{equation}
which is valid only for $\phi>0$. We will discuss $\phi<0$ later, but now we first focus on the $\phi>0$ condition. If we consider the minimal two-field mode, the mass $M$ is constrained by CMB (see Eq. \eqref{cmb}) and also depends on whether it is Higgs-like inflation or $R^2$-like inflation. Usually, $M$ is taken to be  $\mathcal{O}(10^{-5})M_p$ \cite{he2019violent,he2021occurrence}, which corresponds to the mass of scalaron during the small field value in the pure $R^2$ model.  In a minimal two-field mode, one can check the relation between $\xi$ and $M$ in Fig. \ref{xi-m}. 

When $\tilde{\phi}\lesssim  M_p$, the potential can be approximated by
\begin{equation}\label{potentialquadratic}
    U(\phi)\simeq \frac{1}{2}\tilde{M}^2\phi^2-\frac{1}{2}\sqrt{\frac{2}{3}}\frac{\tilde{M}^2}{M_p}\phi^3,
\end{equation}
also automatically giving
\begin{equation}
    \frac{\partial U}{\partial \phi}\simeq \tilde{M}^2\phi-\sqrt{\frac{3}{2}}\frac{\tilde{M}^2}{M_p}\phi^2.
\end{equation}
Here we named this regime to be \textit{quadratic regime}, corresponding to the first term in Eq. \eqref{potentialquadratic} from which the potential is described dominantly by $\phi^2$ term. Also, if we consider the inflaton's trajectory is following \eqref{eominflaton}, the analytical solution is
\begin{equation}
    \phi(t)\simeq \tilde{\phi} e^{-3Ht/2} \cos\left(\left[\tilde{M}^2-\frac{1}{4}(3H)^2\right]^{1/2}t\right),
\end{equation}
with $\tilde{\phi}$ referring to the amplitude of the inflaton during the beginning of the preheating stage. With $\tilde{M}\gg H$, we can claim that the inflaton's oscillation is light-damped during this stage.

If we decompose $\phi$ in Heisenberg representation, namely
\begin{equation}\label{heisenberg}
    \phi(x,t)=\frac{1}{(2\pi)^{3/2}}\int d^3k\left(\hat{a}_k \phi_k(t)e^{-i\bar{k}\cdot \bar{x}} +\hat{a}^\dagger_k \phi^*_k(t)e^{i\bar{k}\cdot \bar{x}}\right),
\end{equation}
we can write the equation of motion for inflaton self-production by
\begin{equation}\label{inflatonself}
    \Ddot{\phi}_k+3H\Dot{\phi}_k+\left(\frac{k^2}{a^2}+\tilde{M}^2-\sqrt{\frac{3}{2}}\frac{\tilde{M}^2}{M_p}\phi\right)\phi_k=0.
\end{equation}
If we rescaled and redefined 
\begin{equation}
    {\varphi_k}\equiv a^{3/2}\phi_k, \hspace{0.5cm} {\kappa^2_\phi}\equiv \frac{k^2}{a^2}+\tilde{M}^2, \hspace{0.5cm} \phi\equiv \tilde{\phi} \sin (\tilde{M} t), \hspace{1cm} \tilde{M}t=2z_\phi+\frac{\pi}{2},
\end{equation}
we can obtain
\begin{equation}
    \frac{d^2\varphi_k}{dz^2_\phi}+\left(A_\phi-2q_\phi\cos(2z_\phi) \right)\varphi_k=0,
\end{equation}
where
\begin{equation}
    A_\phi\equiv \frac{4}{\tilde{M}^2}\left(\frac{k^2}{a^2}+\tilde{M}^2\right) \hspace{1cm}\text{and}\hspace{1cm} q_\phi\equiv 2\sqrt{\frac{3}{2}}\frac{\tilde{\phi}}{M_p},
\end{equation}
found to be the Mathieu equation. $A_\phi \gg q_\phi$ for {the first oscillation after the end of inflation} and $q_\phi$ is much lower just after that, resulting in the resonance getting narrower.
The self-production of the scalaron field belongs to the narrow resonance. As usually happens, the particle fluctuation during the first several oscillations is a rather broad-like  resonance\footnote{see Ref. \cite{kofman1994reheating,shtanov1995universe}}, the resonance self-production of the inflaton field and its decay is neglected. 
With these results, we can assume that the decay product to other channels could be dominant compared with the inflation self-production. One can also realize that the single-field approximation causes $\partial U/\partial h$ to be negligible in Eq. \eqref{hdotdot}, this is resulting in the oscillation being stagnant, which means there is almost no Higgs production due to this mode.

\begin{figure}[t]
\centering
\includegraphics[width=10cm]{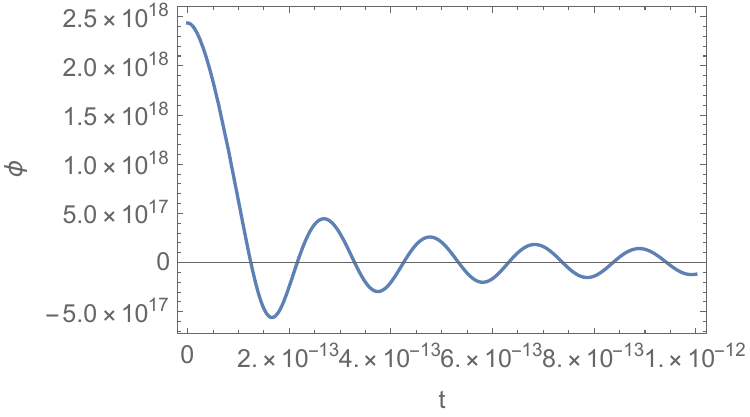}
\caption{Scalaron's time evolution after the end of inflation. The $\phi$ unit is in GeV and $t$ in seconds.}
\label{scalaron}
\end{figure}

We can approximate the potential energy at the end of inflation by imposing the condition $\epsilon = 1$. This yields
\begin{equation}\label{energyend}
    U(\phi_{\text{end}}) < 3.7 \times 10^{-11} M_p^4,
\end{equation}
which is independent of $\xi$, as evaluated using Eq.~\eqref{tildem}. Taking this potential as the initial value for preheating, however, is somewhat subtle. If the scalaron field value at the end of inflation is $\phi_{\text{end}} \sim M_p$, its amplitude decreases sharply after the first crossing. As shown numerically in Fig.~\ref{scalaron}, the field loses about $70\%$ of its value after the first crossing, though the oscillations become smoother thereafter. This suggests the existence of a transition phase between the end of inflation and the onset of preheating, occurring shortly before the first crossing.
For a rough estimate, we therefore take $\phi_{\text{pre}} \sim 0.5 M_p$ as the starting value for preheating, corresponding to the potential
\begin{equation}\label{energypreheating}
    U(\phi_{\text{pre}}) \sim 1 \times 10^{-11} M_p^4.
\end{equation}
We expect the preheating process to deplete most of this energy before the onset of perturbative reheating.

\subsection{The gauge bosons and fermions production}\label{gaugequadratic}
\begin{figure}[t]
\centering
\includegraphics[width=10cm]{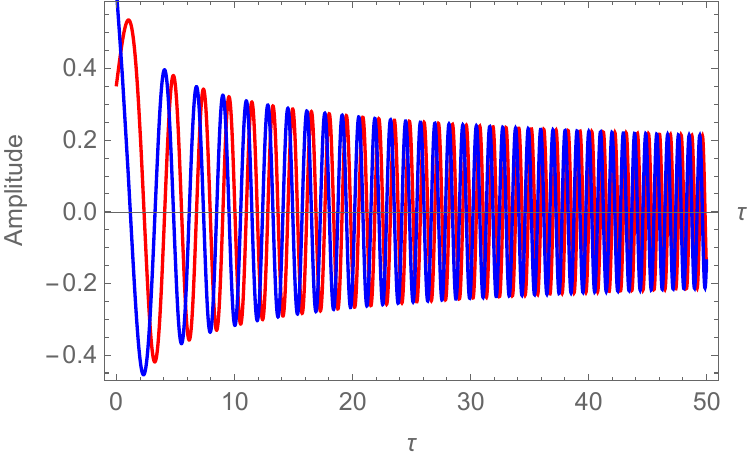}
\caption{Plot of $Ai(\tau_W)$ (\textcolor{red}{red} color) and $Bi(\tau_W)$ (\textcolor{blue}{blue} color). The figure is only up to $\tau_W=50$ to make it clear between red and blue plots. Instead, the process continues until $\sim 10^5$ oscillations (for $\xi\sim 10^2-10^3$), which means until the end of the quadratic regime. The amplitude is in the order of $M_p$.}
\label{Airy}
\end{figure}

As discussed in the previous subsection, the scalaron and Higgs self-production are suppressed. In this case, the production of other fields is expected to be large. Hence, we straightforwardly write the $W$-boson equation of motion by assuming that gauge bosons are scalar fields \cite{bezrukov2009initial}. We assume in this quadratic regime, both $W$ and $Z$ bosons are identical and have the same coupling $g_W$. Hence, we can write the $W$-boson's equation of motion as
\begin{equation}\label{ww}
    \Ddot{W}_k+3H\Dot{W}_k+\left(\frac{k^2_W}{a^2}+{m}_W^2(t) \right)W_k=0, 
\end{equation}
where $m_W^2$ can be defined by (see Eq. \eqref{higgsend})
\begin{equation}\label{mw}
    {m}_W^2=\frac{g^2_W}{4}\sqrt{\frac{2}{3}}\mathcal{C}\frac{\xi M_p}{\lambda}\tilde{\phi}\sin(\tilde{M}t).
\end{equation} 
Please note that the induced mass of $m_W$ was much larger than $\tilde{M}$ at the beginning of the preheating era.
Indeed, the allowed time ($\delta t$) during the zero crossing in which gauge bosons are produced can be written by
\begin{equation}
   \delta t\lesssim \frac{4}{g^2_W}\sqrt{\frac{3}{2}}\frac{\lambda\tilde{M}}{\mathcal{C}\xi M_p \tilde{\phi}}\sim 1\times 10^{-7}/\tilde{\phi},
\end{equation}
which is obtained by the conditions $m_W<\tilde{M}$.
The mass of gauge bosons depends strictly on the amplitude  $\tilde{\phi}(t)$. With those in mind, we approximate $\sin (\tilde{M}t)\simeq \tilde{M}t$. In addition, we can redefine  Eq. \eqref{ww} to be

\begin{equation}\label{wdefine}
    \mathcal{W}_k\equiv a^{3/2}W_k, \hspace{0.5cm} {\kappa^2_W}\equiv \frac{k^2_W}{K^2_W a^2}, \hspace{0.5cm} \tau_W=K_Wt, \hspace{0.5cm}K_W\equiv\left[\frac{g^2}{4}\sqrt{\frac{2}{3}}\mathcal{C}\frac{\xi M_p\tilde{M}}{\lambda}\tilde{\phi} \right]^{1/3},
\end{equation}
thus we obtain
\begin{equation}
    \frac{d^2\mathcal{W}_k}{d\tau^2_W}+\left( \kappa_W^2+\tau_W\right)\mathcal{W}_k=0,
\end{equation}
which belongs to the Airy function. This way, we obtain the solutions
\begin{equation}
    \begin{split}
    &Ai(\tau_W) =\frac{1}{3}i\sqrt{\tau_W}\left[J_{-1/3}(b)+J_{+1/3}(b) \right],\\
    &Bi(\tau_W)=i\sqrt{\frac{\tau_W}{3}}\left[J_{-1/3}(b)-J_{+1/3}(b) \right],
    \end{split}
\end{equation}
with $b=\frac{2}{3}\tau^{3/2}_W$ and $J_{\pm}(\tau_W)$ is the Bessel function of the order $\pm 1/3$. Both functions will run from conformal time $\tau_{W \hspace{1mm} \text{end}}=0$, as we taking the time when inflation is ended by $t=0$, into 
\begin{equation}
    \tau_{W \hspace{1mm}\text{crit}}=K_\text{crit}t_\text{crit}=\left[\frac{g^2}{4}\sqrt{\frac{2}{3}}\mathcal{C}\frac{\xi M_p\tilde{M}}{\lambda}\tilde{\phi}_\text{crit} \right]^{1/3} t_\text{crit}\simeq 4.2\times 10^5 \xi^{-1/3}.
\end{equation}
Taking this result into a plot, we can check it in Fig. \ref{Airy}. In the figure, we only take $\tau_W=50$ since it is more convenient\footnote{Taking $\tau_W \sim 10^5$ may disrupt the figure of both $Ai$ and $Bi$.}. Also, we can predict in $\tau$ more than $20$, the amplitude is only slightly getting smaller, and the oscillation is still not terminated in the quadratic regime. We can approximate the largest $W$-boson production for the first crossing:
\begin{equation}\label{gaugedelltaw}
   \delta {\rho}_W= \int^\infty_0\frac{d k_W^3}{(2\pi)^3} \sqrt{k_W^2/a^2 +m_W^2}e^{-\pi\left(\frac{k_W}{K} \right)^2}\simeq m_W\frac{K^3_W}{8 \pi^3}.
\end{equation}
This way, we obtain the largest energy drain for the first crossing to be 
\begin{equation}
    \delta\rho_W\simeq 4.52 \times 10^{-20} \xi^{3/2} M_p^4.
\end{equation}
 Here we assumed  $k_W^2\ll m^2_W$ in our calculation, and $m_W$ is evaluated along the half oscillation $\int^\pi_0\sin(\tilde{M}t)=1$. Also we used $\tilde{\phi}=\tilde{\phi}_\text{pre}=0.5M_p$.

In this calculation, we confirm that gauge boson production is inefficient in the quadratic regime. The produced gauge bosons continue to oscillate, generating additional daughter fields through secondary parametric resonances, thereby filling the universe with heavy matter during this stage. In contrast, the Fermion production is strongly suppressed by the Pauli exclusion principle. It is consistent with Refs. \cite{risdianto2025second,risdianto2025preheating}. We therefore restrict our attention to gauge bosons. 

By contrast, in Higgs-driven inflation \cite{bezrukov2009initial}, the reheating temperature is determined by gauge boson decay into fermions during the quadratic regime. However, in the present work, we also assume that perturbative fermion production remains heavily suppressed, implying that instant reheating is forbidden in this model. Furthermore, Higgs boson production through oscillating gauge bosons is negligible, as it is suppressed by the factor $\lambda / g_W^2$. 

Finally, we note that tachyonic preheating is in principle possible \cite{he2021occurrence,bezrukov2020heatwave}. Nevertheless, under the assumption that the inflaton trajectory follows the two-field mode effectively as a single-field mode, the tachyonic channel is suppressed due to the smallness of the Higgs field contribution ($\partial U^2/\partial h^2 \simeq 0$) in the quadratic regime.

The inflaton's energy can be approximated to be drained by two gauge bosons by multiplying the series of amplitudes and the number of oscillations. The number of oscillations in the quadratic regime can be approximated by calculating the time started from the beginning of the preheating stage $t_\text{end}$ until critical time $t_\text{crit}$ (which indicates the beginning of the upcoming quartic regime\footnote{See next section about this definition}) as

\begin{equation*}
    t_\text{crit}-t_\text{end}\simeq  \frac{2}{\tilde{M}\mathcal{C}}\frac{\lambda}{\xi }.
\end{equation*}
With the period of oscillation can be approximated by $T=\frac{2\pi}{\tilde{M}}$,  the number of oscillations during the quadratic regime is approximated to be
\begin{equation}\label{noscillation}
    n_\text{osc}=\frac{t_\text{crit}-t_\text{end}}{T}=\frac{1}{\pi \mathcal{C}}\frac{\lambda}{\xi}.
\end{equation}
Thus, we can estimate the total energy drained by the gauge bosons  by using the formula

\begin{equation}
    \rho_\text{gauge}\approx \text{ number of gauge bosons} \times \text{number of crossing}\times\frac{1}{3}\left( \delta{\rho}_W \times n_\text{osc} \right).
\end{equation}
The $\left( \delta{\rho}_W \times n_\text{osc} \right)$ should be proportionally resemble Fig. \ref{Airy} for $\tau=\infty$. Thus, it is where the factor $1/3$ numerically comes from. Finally we get
\begin{equation}\label{gaugeenergy}
    \rho_\text{gauge}\sim  6 \times 10^{-14}\sqrt{\xi} M_p^4.
\end{equation}
In Eq. \eqref{gaugeenergy}, if the non-minimal coupling is in $\mathcal{O}(10^4)$, they could drain the whole inflaton's energy. However, for the maximum value of $\xi$ in this model ($\xi\sim 4500$), it still can not drain the whole inflaton's energy. In this case, we proposed a new field with a larger coupling than gauge bosons coupled with the Higgs. We will discuss this possibility in the next subsection.

\subsection{The wear off gravitational effect, end of quadratic regime, and dark matter}
The universe is filled with nothing but inflaton during inflation, leaving most of it nearly a vacuum. As we consider the second Friedman equation
\begin{equation}
    \Ddot{a}=-\frac{1}{6 M_p^2}(\rho_\phi+3P_\phi)a,
\end{equation}
the pressure $P_\phi=-\rho_\phi$ causes acceleration. This condition happens until the kinetic term of scalaron $\propto \Dot{\phi}^2$ gets large and comparable with the inflaton's potential. The cosmological constant is only responsible for the expansion in the post-inflationary era, which we omitted in the last equation. At this point, the kinetic term of Higgs is still suppressed by $e^{-\chi}$ and tends to be neglected. Thus, the scalaron remains dominant at this point. This condition happens simultaneously with the creation of massive fields, which constantly fill the universe, leaving the universe to enter the (heavy) matter-dominated era. At the end of the quadratic regime, when the field value of the scalaron enters the critical point, the gravitational effect is wearing off. It is shown when $\tilde{\phi}_\text{crit}\approx \tilde{h}_\text{crit}$. This implies that Higgs, as the subdominant component of the inflaton, now becomes dominant. The details of this process can be followed in the section \ref{quartic}.

In this subsection, we will introduce a Lagrangian of the dark matter (DM), which we assume is only coupled with the Higgs and invariant under global $Z_2$ symmetry. The importance of this DM is due to the fact that both gauge bosons failed to drain the inflaton's energy. The $S$ field is considered the DM candidate in this model.\footnote{Since DM is known to be more abundant than SM particles, we argue that the DM with a stronger coupling compared to SM particles could effectively drain the inflaton's energy.} Here we have $S=(0 \hspace{0.5cm} s)^\top/\sqrt{2}$ and $\braket{S}=0$. The additional potential from Eq. \eqref{h-r2action} with $S$-field  can be written by\footnote{Here we assumed the term  $\frac{1}{2}m_s^2 s^2$ and $\frac{1}{4}\lambda_s s^4$ are much smaller during this time.}

\begin{equation}\label{lagrangedm}
    \frac{1}{4}\lambda_{hs} h^2 s^2.
\end{equation}
During the quadratic regime, Eq. \eqref{lagrangedm} will be transformed by the Weyl transformation on the Einstein frame into
\begin{equation}\label{smass}
    \frac{1}{2\sqrt{{6}}}\mathcal{C}\frac{\xi M_p}{\lambda}\lambda_{hs} \phi \hspace{1mm} s^2.
\end{equation}
This means that during the quadratic regime, DM production can be produced by the perturbative decay of the scalaron at the tree level, which also plays the dominant role in the decay of the scalaron. However, this effect should never be significant during the preheating stage, as the parametric resonance of DM takes over. Before we proceed, we assume  $\lambda_{hs}$ value\footnote{For simplicity, we assumed that the additional fields (the DM) would not affect the Higgs' running coupling and ruined our setting of  $\lambda=0.01$.} is $\lambda< g^2< \lambda_{hs}\simeq 1$. With the large coupling of $\lambda_{hs}$, we expect both parametric resonance and perturbative decay of scalaron to DM to be dominant compared to the gauge bosons and fermions productions and successfully drain the inflaton's energy. 

The decay rate of scalaron to two DMs can be obtained as
\begin{equation}\label{inflatondecaydarkmatter}
    \Gamma_{\phi\rightarrow ss}=\frac{1}{384\pi}\mathcal{C}^2\frac{\xi^2 M_p^2}{\lambda^2 \tilde{M}}\lambda_{hs}^2.
\end{equation} 
As we continue, 
using our previous constraint, the decay should be much lower than the Hubble parameter ($\Gamma_{\phi\rightarrow ss}\ll H$). The DM density only fills a minor part of the universe during the early preheating stage, but grows significantly due to parametric resonance. In addition, it is truly remarkable that during preheating, particle production by perturbative decay should be much smaller than the resonance production for successful preheating. 

However, when the parametric resonance is extremely narrow, the resulting particle production due to parametric resonance is suppressed at the end of the quadratic regime. The perturbative effect, enhanced by the Bose-Einstein condensation (BEC),  drains the inflaton's energy. This enhanced\footnote{Let us call it that way, which is a perturbative effect enhanced by BEC.} perturbative effect, even though it is small, is constantly draining the inflaton field. This effect becomes important during the transition of the quadratic and quartic regimes\footnote{This will be defined shortly in the next section}. Nevertheless, we could not confirm this condition in the numerical study, but we assumed such a condition existed during the transition. The reason is that at the end of the quadratic regime, the oscillation becomes extremely narrow, and a mechanism should exist to drain the inflaton's energy. This is to ensure the inflaton's field value drops at a certain favored point, e.g. $\tilde{\phi}_\text{crit}$.

The decay rate enhanced by BEC can be written as \cite{mukhanov2005physical}
\begin{equation}\label{decays}
    \Gamma_\text{eff}\simeq\Gamma_{\phi\rightarrow ss} (1+2\bold{n}_k^{s}).
\end{equation}
We assume the initial condition of occupation number of scalaron found to be much larger than DM $\bold{n}_k^s$.
In addition, the equation of motion in momentum space of DM coupled with scalaron can be written by
\begin{equation}\label{dmeom}
    \Ddot{s}_k+\left(\frac{k_s^2}{a^2}+\frac{1}{\sqrt{6}}\mathcal{C}\frac{\xi M_p}{\lambda}\lambda_{hs} \tilde{\phi}\sin(\tilde{M}t)\right)s_k=0.
\end{equation}
The width of $k_s$, which corresponds to the width band of the created particles, can be evaluated by assuming the energy of a single $S$ equals $\tilde{M}/2$, and we obtain
\begin{equation}
    \left(\frac{\tilde{M}}{2}\right)^2=\frac{k^2_s}{a^2}+\frac{1}{\sqrt{6}}\mathcal{C}\frac{\xi M_p}{\lambda}\lambda_{hs} {\phi}.
\end{equation}
Finally, $\Delta k_s$ can be evaluated as
\begin{equation}
    \Delta k_s =\sqrt{\frac{2}{3}}\mathcal{C}\frac{\xi M_p}{\lambda \tilde{M}}\lambda_{hs}\tilde{\phi},
\end{equation}
where we neglected the expansion of the universe. We can calculate the occupation number $\bold{n}_k^s$ as
\begin{equation}\label{nk}
    \bold{n}_k^s=\frac{n_s}{4\pi k_s^{*2}\Delta k_s/(2\pi)^3}=\sqrt{\frac{3}{8}}\frac{ \tilde{\phi}}{\pi \mathcal{C} \xi M_p }\frac{\lambda}{\lambda_{hs}}\frac{n_s}{n_\phi},
\end{equation}
where we have evaluated Eq. \eqref{nk} by using $k_s^*=\tilde{M}/2$ and $n_\phi=\frac{1}{2}\tilde{M}\tilde{\phi}^2$.
For further usage, it is important to write Eq. \eqref{dmeom} into the redefined form as
\begin{equation}\label{sdefine}
    \mathcal{S}_k\equiv a^{3/2}s_k, \hspace{0.5cm} {\kappa^2_s}\equiv \frac{k_s^2}{K_s^2a^2}, \hspace{0.5cm} \tau_s=K_st, \hspace{0.5cm}K_s\equiv\left[\frac{\lambda_{hs}}{\sqrt{6}}\mathcal{C}\frac{\xi M_p\tilde{M}}{\lambda}\tilde{\phi} \right]^{1/3},
\end{equation}
thus we obtain
\begin{equation}\label{lames}
    \frac{d^2\mathcal{S}_k}{d\tau^2_s}+\left( \kappa_s^2+\tau_s\right)\mathcal{S}_k=0,
\end{equation}
which also belongs to the Airy functions. The parametric resonance from this equation can enhance the BEC on this perturbative decay. Thus, the particle number density of the DM is calculated via
\begin{equation}
     n_s=\int^\infty_0 \frac{d^3\kappa_s}{(2\pi)^3}e^{-\pi \frac{\kappa^2_s}{K_s^2}}=\frac{1}{8\pi^3}K_s^{3}.
\end{equation}
We can substitute the last result into Eq. \eqref{nk} and finally into \eqref{decays} to obtain the effective decay of $\Gamma_{\phi\rightarrow ss}$.

Before we proceed, if the enhanced decay rate at the end of the quadratic regime is taken into account, which is $\Gamma_\text{eff}$ (see Eq. \eqref{decays}), we can compare its result to the Hubble parameter $H$ during the same period. The end of preheating can be evaluated by using the relation $\Gamma_\text{eff}\gtrsim H$. The Hubble parameter can be obtained by using the relation
\begin{equation}
    H^2=\frac{\rho_\phi}{3M_p^2},
\end{equation}
where $\rho_\phi$ is evaluated by the energy density during the near end of the quadratic regime, we show that for the preheating to be followed by the quartic regime, the non-minimal coupling should be constrained as
\begin{equation}
    \xi\lesssim 4.2, 
\end{equation}
where we have calculated the BEC effect with $n_s$ in Eq. \eqref{nk} from which is evaluated using the idea that DM density in the late quadratic regime grows as much as scalaron during the beginning of the quadratic regime, thus $n_s/n_\phi\sim 1$.

To calculate the DM particle production due to parametric resonance, we can use Eq. \eqref{lames}. The solution of this equation also belongs to the Airy function. Furthermore,
the energy density for each crossing can be calculated similarly with Eq. \eqref{gaugedelltaw} as
\begin{equation}\label{gaugedelltas}
   \delta {\rho}_s= \int^\infty_0\frac{d k_s^3}{(2\pi)^3} \sqrt{k_s^2/a^2 +m_s^2}e^{-\pi\left(\frac{k_s}{K_s} \right)^2}\simeq m_s\frac{K_s^3}{8 \pi^3},
\end{equation}
where the induced mass $m_s$ is evaluated from Eq. \eqref{smass}.
Finally, the total energy drain by DM can be calculated similarly to Eq. \eqref{gaugeenergy} by
\begin{equation}
    \rho_{s}\sim  6 \times 10^{-13}\sqrt{\xi}M_p^4.
\end{equation}
Interestingly, only if $\xi\gtrsim 277$ could the whole inflaton's energy be drained. It means that if we used a smaller non-minimal coupling, the inflaton's energy could not be drained by resonance alone. This way, the remaining inflaton's energy will be drained by perturbative decay. In addition, it will affect the reheating temperature. We will discuss this matter further in the reheating section \ref{reheating}.

The mass of DM can be constrained by using the DM abundance $\Omega h^2=0.12$ \cite{aghanim2020planck}. Straightforwardly, we obtain\footnote{The details of this calculation can be found in Ref.\cite{tanedo2011defense}} 
\begin{equation}
    \Omega h^2 = \frac{1.55\times 10^{-10}}{\braket{\sigma_{ss\rightarrow hh} \nu}}\text{GeV}^{-2}.
\end{equation}
With $\braket{\sigma_{{ss\rightarrow hh}} \nu}$ corresponding to the annihilation of the DM to 2 Higgs bosons, we can constrain the DM mass to be $m_s\sim 500$ GeV. We expect the DM to be a value that can be detected in the near future, such as in the Large Hadron Collider (LHC) or International Linear Collider (ILC).

\section{Preheating in the Quartic Regime}\label{quartic}
When the inflaton field reaches a critical value $\tilde{\phi}_\text{crit}$, preheating in the quadratic regime comes to an end. The potential then takes the form $\frac{1}{4}\lambda \phi^4_\text{crit}$. We note that even though it may appear as if we are changing frames (from Einstein to Jordan), in principle, we are still working within the Einstein frame. This follows precisely the mechanism discussed in the Ref. \cite{risdianto2025second}.
Since $\Tilde{\phi}_\text{crit} \simeq \Tilde{h}_\text{crit}$ at this stage, the potential can equivalently be written as 
\begin{equation}
    \frac{1}{4}\lambda h^4.
\end{equation}
We therefore refer to this stage as the \textit{quartic regime}, which corresponds to its new potential. In a physical context, the Higgs boson, which was subdominant during inflation, now takes over as the oscillating inflaton field and continues the preheating process into the quartic regime. During this time, the universe enters the radiation-dominated era. In addition, we can approximate the remaining energy of the  inflaton  to be
\begin{equation}\label{potentialquartic}
    V(h)\simeq  5.7 \times 10^{-33} \xi^4 M_p^4,
\end{equation}
which was found to be $\xi$ dependent.

The equation of motion for Higgs self-production can straightforwardly be written in the Heisenberg representation, analogous to Eq. \eqref{heisenberg}, as

\begin{equation}\label{hequation}
    \Ddot{h}_k+3H\Dot{h}_k+\left(\frac{k^2_h}{a^2}+3{\lambda}h^2 \right) h_k=0.
\end{equation}
Defining
\begin{equation}\label{kappatau}
    \bold{h}_k=a h_k \hspace{0.5cm}
    \kappa^2_h\equiv\frac{k^2_h}{ {\lambda}\tilde{h}^2}, \hspace{0.5cm}a(\tau)=\frac{1}{2\sqrt{3}}\frac{\tilde{h}}{M_p}\tau,\hspace{0.5cm} \tau\equiv\left(6 {\lambda}M_p^2/\pi\right)^{1/4}\sqrt{t},
\end{equation}
we can write Eq. \eqref{hequation} in the more straightforward as
\begin{equation}\label{eomquartich}
    {\bold{h}_k}''+\left(\kappa^2_h+3cn^2\left(\tau, \frac{1}{\sqrt{2}}\right) \right)\bold{h}_k=0, \hspace{0.5cm},
\end{equation}
where $\tilde{h}$ corresponds to the amplitude of the Higgs field, and the prime corresponds to the derivative with respect to conformal time $\tau$. Also, the solution of $\bold{h}_k(\tau)=\bar{\bold{h}}_k f(\tau)$ is obtained in the same way as Ref. \cite{greene1997structure}:
\begin{equation}
    f(\tau)=cn\left(\tau, \frac{1}{\sqrt{2}} \right),
\end{equation}
which is an elliptic cosine function.

In the same manner, we examine the production of gauge bosons and fermions during this period. In the quartic regime, Pauli blocking is fully lifted, and fermion production through both parametric resonance and perturbative decay is no longer suppressed. For simplicity, we treat fermions effectively as bosons,\footnote{While the resonance structure for fermions differs from that of bosons, they can be approximated as bosons to a certain degree of accuracy \cite{greene2000theoryfermion}.} which allows us to apply the same analytical treatment.

To get the particle production, it is important to start with the equation of motion of the $W$-boson in a similar way to Eq. \eqref{ww} in the Heisenberg picture as
\begin{equation}\label{wwquartic}
    \Ddot{W}_k+3H\Dot{W}_k+\left(\frac{k^2_W}{a^2}+{m}_W^2(t) \right)W_k=0, \hspace{0.5cm} m^2_W=\frac{g^2}{4}h^2
\end{equation}
The analytical calculation of the $W$-boson will be different in this regime compared to the quadratic ones. By following the redefinition in Eq. \eqref{kappatau} and $\mathcal{W}_k=a W_k$, we arrived at
\begin{equation}\label{eomquarticw}
    \mathcal{W}_k''+\left(\kappa^2_W+\frac{g^2}{4 {\lambda}}cn^2\left(\tau, \frac{1}{\sqrt{2}}\right) \right)\mathcal{W}_k=0.
\end{equation}
We can generalize Eq. \eqref{eomquartich} and \eqref{eomquarticw} so they can be used for any species, which is

\begin{equation}\label{eomquarticgeneral}
    {\varphi_k}''+\left(\kappa^2_\varphi+\Upsilon cn^2\left(\tau,\frac{1}{\sqrt{2}} \right) \right)\varphi_k=0,
\end{equation}
where (for instance) $\Upsilon=3$ correspond to the Higgs self-production, $\Upsilon=\frac{g^2}{4\lambda}$ for gauge bosons, $\Upsilon=\frac{y_\psi^2}{2\lambda}$ for fermions, and $\Upsilon=\frac{\lambda_{hs}}{4\lambda}$ for DM.

\begin{figure}[t]
\centering
\includegraphics[width=16cm]{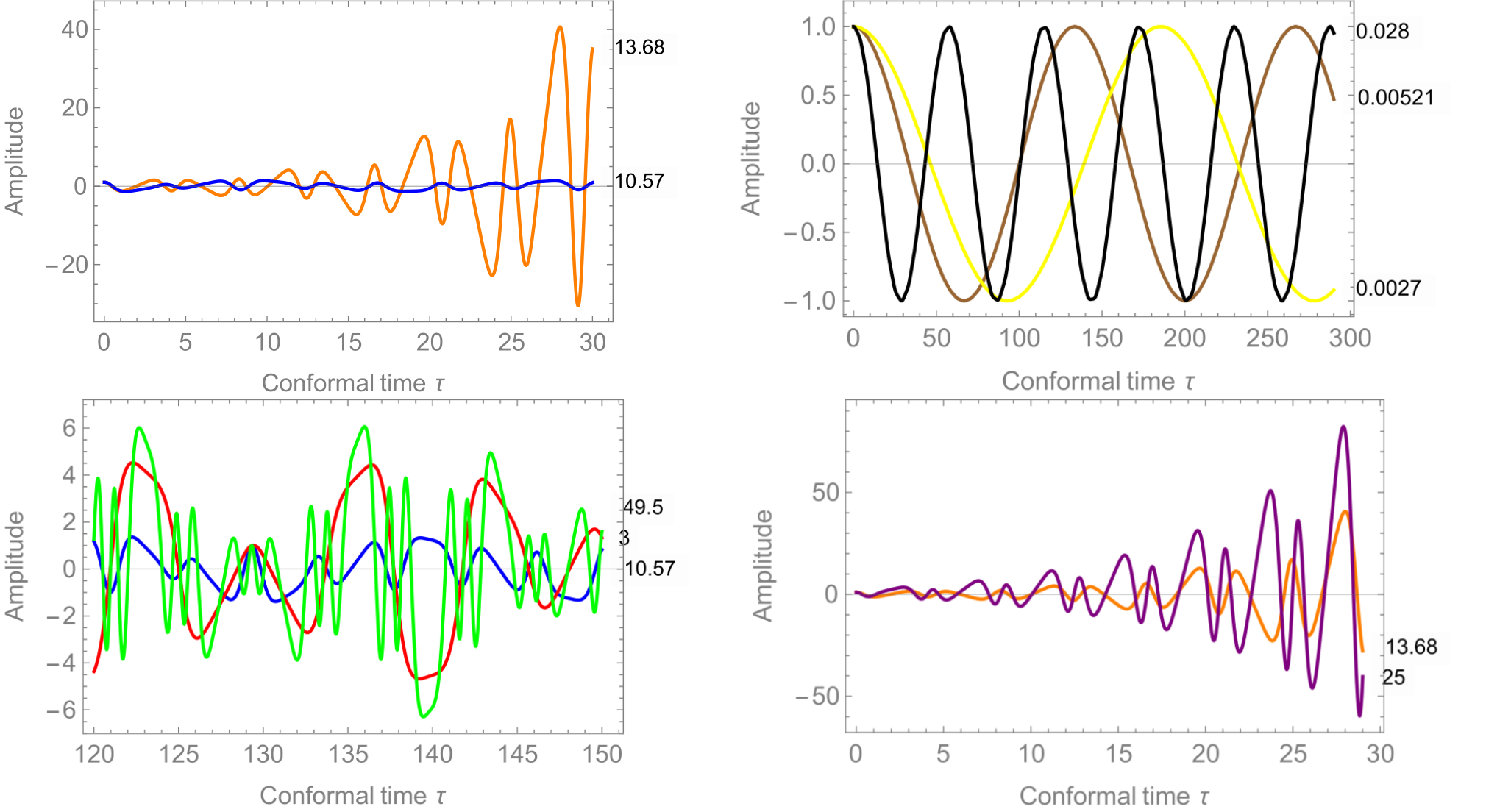}
\caption{Plot of varied $\Upsilon$ which refer to different particle productions. The color difference refers to different particles. \textcolor{blue}{blue}: $W$ boson, \textcolor{orange}{orange}: $Z$ boson, \textcolor{red}{red}: Higgs boson, \textcolor{green}{green}: Top quarks, \textcolor{brown}{brown}: Tauon, \textcolor{black}{black}: Bottom quarks, \textcolor{yellow}{yellow}: Charm quarks, and \textcolor{purple}{purple}: the DM ($S$-field). The values on the right of every plot correspond to the values of $\Upsilon$.}
\label{numericquartic}
\end{figure}

In the numerical analysis of Eq.~\eqref{eomquarticgeneral}, we vary $\Upsilon$ according to several Standard Model (SM) particles, while assuming $\kappa^2=0$. In particular, we test $\Upsilon$ for SM particles as well as for the dark matter (DM) candidate, denoted by the $S$-field. In our earlier analysis, we treated the $W$ and $Z$ bosons as identical for simplicity. However, in the numerical results presented in Fig. \ref{numericquartic} (upper-left panel), we explicitly distinguish their couplings\footnote{The couplings of the SM particles can be seen in Particle Data Book \cite{navas2024review}}. We find that even a small difference in couplings leads to significantly different production rates between the $W$ and $Z$ bosons.  

An additional feature arises for $\Upsilon \simeq 1\text{--}2$, where particle production exceeds that of the $Z$ boson. Since no known particle in our setup has such couplings, we do not discuss this case further. Conversely, we also examine particles with small couplings (small $\Upsilon$), shown in Fig. \ref{numericquartic} (upper-right panel), whose production can be safely neglected. Furthermore, we analyze the couplings of heavy quarks (Fig. \ref{numericquartic}, bottom-left panel), which yield only a modest enhancement in particle production. Finally, in Fig. \ref{numericquartic} (bottom-right panel), we compare the DM production with that of the $Z$ boson.

The variation of $\kappa_\varphi$ can either enhance or suppress particle production. The conditions under which this occurs are discussed in Ref. \cite{greene1997structure}, where the corresponding instability chart is also provided. In the present work, however, we do not explore such behavior.  

\section{The Reheating Temperature}\label{reheating}

The discussion of inflation would be incomplete without addressing the reheating temperature. To this end, we revisit the preheating dynamics in the quadratic regime. Due to the Pauli exclusion principle, we assume that instantaneous reheating through the decay of gauge bosons into lighter fermions is suppressed. Consequently, for $\xi < 4.2$, the $Z$ bosons produced through inflaton resonance in the quartic regime are expected to play the primary role in generating the reheating temperature via their decay channels.\footnote{One should note that if $\xi > 4.2$, the preheating is stopped during the quadratic regime.}
In this work, we focus on the dominant decay mode $Z \rightarrow \bar{b}b$. 

The conformal mass of the $Z$ boson and its number density, respectively, can be written as
\begin{equation}
    \tilde{m}_W=\sqrt{\frac{g^2_W}{4\lambda}}f_0 \sin(c f_0 \tau) \hspace{0.5cm} \text{and}\hspace{0.5cm} \bar{n}_W=\frac{1}{8\pi^3}\left( \frac{g^2_W}{4\lambda}\right)^{3/2},
\end{equation}
where $c=\frac{2\pi}{\tau_0}$ and $\tau_0=7.416$ \cite{greene1997structure} and we assumed there is no change in $f_0$. Again, we assumed the $W$ and $Z$ bosons couplings and masses are indistinguishable to simplify the calculation\footnote{That is why we still used $\bar{m}_W$ and $\bar{n}_W$ instead of $\bar{m}_Z$ and $\bar{n}_Z$. }. The energy density of the produced $Z$ boson in the first half-crossing can be approximated as \cite{ballesteros2017standard}
\begin{equation}\label{firsthalfcrossingdensity}
    \delta \bar{\rho}_Z =\int^{\tau_0/2}_0d\tau \bar{\Gamma}_{Z\rightarrow \bar{b}b} \bar{m}_W \bar{n}_W e^{\int^\tau_0 \bar{\Gamma}_{Z\rightarrow\bar{b}b}d\tau'},
\end{equation}
where $\bar{\Gamma}_{Z\rightarrow \bar{b}b}$ correspond to the decay of $Z$ boson in the conformal mode. The energy density transferred to the light particles is averaged and estimated as
\begin{equation}
    \bar{\rho}_Z=\frac{2\delta\bar{\rho}_Z}{\tau_0}\tau\simeq 0.04\tau.
\end{equation}
If the inflaton's oscillation energy density in conformal mode is $\bar{\rho}_h=\frac{1}{4\lambda}$ and $\bar{\rho}_Z$ are conserved, means $\bar{\rho}_h=\bar{\rho}_Z$, it can be translated into physical unit $\rho_h$ as
\begin{equation}
   {\rho}_h=\frac{1}{4\lambda}\left(\frac{\sqrt{\lambda}\tilde{h}_\text{crit}}{a} \right)^4=\frac{\pi^2 g^*}{30}T_R^4,
\end{equation}
where $g^*\sim 100$ corresponds to the SM particle's degrees of freedom. Using Eq. \eqref{kappatau}, we can obtain the reheating temperature $T_R$ as
\begin{equation}\label{temp}
    T_{R(\xi<4.2)}\sim 1\times 10^{15}\text{GeV}.
\end{equation}

It should be noted that at $\tilde{h}_\text{crit}$ the total energy density of the inflaton is given by Eq. \eqref{potentialquartic}. However, the energy transferred to $Z$ bosons during a single half-oscillation, $\rho_Z$, exceeds the residual inflaton energy described in Eq. \eqref{potentialquartic}. This implies that the inflaton energy can be efficiently drained by a single zero-crossing of the inflaton field. Accordingly, the reheating temperature in Eq. \eqref{temp} must be re-evaluated using the total remaining inflaton energy of Eq.~\eqref{potentialquartic}. From this procedure, we obtain the following upper bound on the reheating temperature:
\begin{equation}\label{tr-upper}
    T_{R(\xi<4.2)} \simeq 1.17 \times 10^{10}\,\xi \;\text{GeV}.
\end{equation}
Since $\xi$ lies in the interval $1 < \xi < 4.2$, the reheating temperature remains within the order of $10^{10}$ GeV. This mechanism differs from that in Higgs inflation \cite{bezrukov2009initial}, where gauge boson decays occur instantaneously after their production in the quadratic regime. In contrast, in our case, the decay is delayed until the quartic regime, such that the total energy transfer per crossing is bounded by the inflaton energy at that time\footnote{An alternative treatment of the reheating temperature within the same model, but restricted purely to the quadratic regime, is given in Ref. \cite{he2021perturbative}.}.

In the other case, for $\xi \gtrsim 4.2$, the inflaton oscillations terminate before entering the quartic regime. During this period, gauge bosons—subdominant only to dark matter—decay into relativistic particles and thus provide the main contribution to the reheating temperature. In this regime, the reheating temperature can be approximated as $T_{R(\xi>4.2)} \approx 0.55 \sqrt{M_p \Gamma_W}$, where $\Gamma_W$ denotes the total decay width of the gauge bosons. Consequently, the reheating temperature is estimated to be
\begin{equation}
    T_{R(\xi>4.2)} \sim 1 \times 10^9 \, \text{GeV}.
\end{equation}

\section{Conclusion}
We have investigated Higgs-$R^2$ inflation, focusing on both its inflationary and reheating features. In this work, we employed the effective single-field approximation, namely the minimal two-field mode along the inflaton trajectory. Within this framework, the inflaton evolves along the direction of the effective mode. Our analysis shows that the turn parameter $\eta^{\perp}$ depends strongly on the Higgs parameters, in particular the non-minimal coupling $\xi$. The turn parameter could, in principle, affect the largeness of non-Gaussianity $f_\text{NL}$. However, our analytical approximation indicates that, even for large $\xi$, the resulting contribution is insufficient to generate significant non-Gaussianity. This outcome is consistent with the expectations of an effective single-field model. Consequently, the non-minimal coupling $\xi$ can be treated as a free parameter, as no definitive constraints apply. Therefore, $\xi$ may range from small values, corresponding to a nearly pure $R^2$ model, to large values, corresponding to nearly pure Higgs inflation.

In the preheating stage, we divided it into two regimes: the quadratic regime and the quartic regime. Considering the quadratic regime, we found that the gauge bosons dominate particle production. But, they failed to drain the whole inflaton's energy. Thus, we introduced the DM candidate with a large coupling. We expected this DM candidate could potentially drain the whole inflaton's energy. During this quadratic regime, it strongly depends on $\xi$ to obtain the different conditions in the preheating stage. If the non-minimal coupling $\xi$ is large, which is $\gtrsim 277$, the resonance production of the inflaton could potentially drain the whole inflaton's energy density. On the contrary, if the non-minimal coupling $\xi$ is less than $4.2$, the resonance continues to the quartic regime. For the value in between ($4.2 <\xi <277$), the inflaton energy is partly converted to heavy particles via resonance.

We have evaluated the reheating temperature across three distinct ranges of the non-minimal coupling, focusing on the associated reheating features. Since fermion production is suppressed during the quadratic regime, both resonant production and perturbative decay into fermions are strongly disfavored. As a result, the reheating temperature determined via the decay of gauge bosons into fermions becomes relevant only at the end of the quadratic regime. For $\xi>277$, the reheating temperature is set by the decay of gauge bosons produced resonantly into fermions at the end of this regime. In the intermediate range $4.2<\xi<277$, the remaining inflaton energy is existed; however, due to the dominance of gauge bosons throughout the quadratic regime, the reheating temperature obtained is nearly identical to the case of $\xi>277$. Consequently, for $\xi>4.2$, the reheating temperature is limited to $T_{R(\xi>4.2)} \sim 10^9 \,\text{GeV}$. In contrast, for $\xi<4.2$, the reheating temperature is determined by the decay of $Z$ bosons—produced resonantly in the quartic regime—into fermions. In this case, the reheating temperature is bounded by the remaining inflaton energy at $\tilde{h}_\text{crit}$, yielding a higher value on the order of $\sim 10^{10} \,\text{GeV}$.

Finally, based on the range of $\xi$, the mechanism with $\xi> 4.2$ may lead to a longer preheating duration. Consequently, the inflaton energy transfer is inefficient\footnote{One should have noted that the preheating stage is introduced to efficiently drain the inflaton energy to avoid extremely high reheating temperature.}. The large $\xi$ is suffered by the unitarity issue. In summary, the small $\xi \lesssim 4.2$ is more favorable due to the unitarity issue, the shorter preheating duration, and also the favored reheating mechanism.

\begin{acknowledgments}
It is a pleasure to thank Daijiro Suematsu, Ahsani Hafidzu Shali, and Idham Syah Alam for the helpful discussion. 
\end{acknowledgments}




\nocite{*}

\bibliography{apssamp}

\end{document}